# An Authoring System for Editing Lessons in Phonetic English in SMIL3.0

**Merzougui Ghalia[1] and Djoudi Mahieddine[2]**

**[1] Departement of Informatics, University of Batna, 05000, Algeria**

**[2] Laboratory XLIM-SIC and IRMA a Research Group, UFR Sciences SP2MI, University of Poitiers Teleport 2, Boulevard Marie et Pierre Curie BP 30179 86962 Futuroscope, Chasseneuil Cedex- France**

**Abstract**

One of the difficulties of teaching English is the prosody, including the stress. French learners have difficulties to encode this information about the word because it is irrelevant for them. Therefore, they have difficulty to produce this stress when they speak that language. Studies in this area have concluded that the dual-coding approach (auditory and visual) of a phonetic phenomenon helps a lot to improve its perception and memorization for novice learners. The aim of our work is to provide English teachers with an authoring named SaCoPh for editing multimedia courses that support this approach. This course is based on a template that fits the educational aspects of phonetics, exploiting the features of version 3.0 of the standard SMIL (Synchronized Multimedia Integration Language) for the publication of this course on the web.

## 1. Introduction

The English and French share a large lexicon, where the spelling forms of a word in both languages are similar. However, the accentual system established by the two languages for these words make them opaque for oral learners. It is observed that certain syllables are more readily audible than others. We speak about accented syllables in this case.

The French learner faces two challenges: to perceive, in the listening phase, the accented and unaccented syllables, and reproduce during the production phase, a sufficient contrast between the two types of syllables.

French students may have serious gaps in phonetic and prosodic when making oral presentations, despite a correct language on the lexical and syntactic. The absence of discrimination vowel/diphthong and the displacement of stress make some words unrecognizable.

Controlled empirical studies confirm what teachers observe every day. These studies show that Anglophone Canadians recognize less isolated words pronounced by a French Canadian. The authors attribute this difference to a lack of emphasis. But [6] confirms that this problem is not sensory but lies at the level of working memory; there is negligence when encoding information. French students learning English fail to treat stress, which has little value in their native language, and therefore they do not store this information. During their presentations, they will put the accent on a random syllable of an English word, and this indicates negligence in encoding and a lack of storage of the place of the stress and not deafness or a production problem. This has a negative effect on understanding their speech.

There are some students who after 10 years of learning English language, have not yet mastered the pronunciation of words that seem elementary (like: who, women, chocolate, village, low, allow, sun, sound). Series such that (there're, aren're, were not, were, where) or graphically similar words such as (tough, trough, though, through, thought) pose enormous problems of memorization in oral.

In [7], Beck et al have made the hypothesis that attention processes play a fundamental role in this case. The visual computing solutions seem to be a good solution. Visual tracking helps in distinction of parts of speech in which problems of perception and understanding arise. The sound becomes visually observable in time, unlike its first material form of transient vibrations of air. The pronunciation would be easier if the student simultaneously read and hear the word 'development', where the stressed syllable is underlined visually (a different style, font and color associated). Treating such a word is in auditory encoding of the linguistic information and in a visual encoding of both language and paralinguistic information, hence the need for a tool for editing documents supporting such a presentation.

In section two, we present the tools for pronunciation and the tools for editing multimedia documents in SMIL standard, and we will position our system with respect to these two groups of tools. Section three presents the model of document proposed to support the approach of dual coding of phonetic aspects to teach. Then, in section four we describe the architecture of our authoring. The article ends with a conclusion and some perspectives.

## 2. State of the Art

### 2.1 Computing Tools of Pronunciation

In the recent years and as regards the acoustical phonetic, advances in graphic display screens are spectacular. Software such as (speaker, tell me more, English plus, book or voice) using oscillo-grams to present the voice have a limited supply, while (Win Pitch, Speech analysis), using curves of fundamental frequencies, are too limited in use because of their complexity or because of ergonomics errors.
In [9], the authoring Sound Right was based on the basic curves for drawing simplified intonative curves using extensible arrows that appear below the text. The difficulty of interpreting complex curves explains their limited use in language courses in schools and universities.
Prosodic Font is a system developed at MIT Media Lab [12] to automatically generate dynamic fonts (from an oral speech input) that vary with the time and the change of tone in a speech. The goal is to generate animated text depending on the intonation and on the prosody of speech. Such a solution is not accepted as the best didactic solution for teaching pronunciation.

### 2.2 Publishers of Multimedia Course in SMIL Format

In this section we briefly describe three editors of SMIL documents:
SWANS [7] is an authoring system that allows any teacher to semi automatically generate multimedia documents where the accent is marked visually (by typographical markers such as color style ...) and aurally. The generated document is a web page where the learner can read and/or listen to a speech synchronized with the text annotations. The scenario requires initial import of media (text, audio and video) in the work environment, then, synchronization of the text (which is segmented into units of breath) with its audio or video pronunciation and finally, the teacher can annotate the text by typing markers.
The system LimSee3 [2] is a multimedia publisher of new generation, which uses templates to simplify editing and ease repetitive tasks. It also allows users to generate documents for different output formats (SMIL, XHTML + JavaScript and timesheets). Currently, there are three templates that are based on the construction of a multimedia course:
The first model allows to build a slide show (set of slides) to prepare a course. Each slide can contain one or more media. These media are inserted or imported from outside by simple gestures (copy and paste or drag and drop).
The second template enables his user to build a course record. In this case, the issue needs first to import the slides used during the lesson or the teacher's image and voice (video and audio track), then to synchronize them by adjusting the transparencies with audio and video. The synchronization tool allows replay of the audio plug and indicates by clicks the times when we need to change the transparency.
The third model can annotate in real time, oral examinations of students.
The publisher ECoMaS [8] is also an editor of mediated courses and is based on document’s models. The final presentation is generated in the same way as in the second model LimSee3, but the scenario of edition is different. The editing by ECoMaS requires to import transparencies which are images, then one has to record the oral explanations of each, then the system generates a publishable presentation on the web (SMIL2.0)[11], where slides are synchronized with their audio explanation and an index table which provides temporal navigation during the presentation of the course.

The last three editors use document templates organized hierarchically. Each model is seen as a document with holes, where the user simply fills the holes by media (text, image, audio or video). It is clear that these Medias are imported from external sources and therefore the teacher must prepare in advance each of them with its corresponding tool. This is tedious, especially if it is to edit formatted text (with colors and styles ...) and then associate it with pronunciation; he must use two different tools (one for word processing and one for the sound), import them into the system and then synchronize them.
An annoying limitation of these tools is the lack of a graphics tool to edit formatted text. Versions of SMIL 1.0 and 2.0 used by these tools do not support tags for text formatting (color, style, font ...), which is very important to materialize the dual coding approach mentioned in our problem. SWANS uses standard XHTML + SMIL. ECoMaS uses the language RealText only to make the title of a transparent and the content of the index table.
The latest version of SMIL [4] supports text formatting features within document (.Smil), but so far there is no graphical editor for this version. Our work is the first contribution to the development of such an editor.
Using the International Phonetic Alphabet (IPA), phonetic attempts to represent the sounds more accurately but teachers cannot use this type of character with existing publishers.

## 3. The SaCoPh Approach

### 3.1 Objectives

-Improve the collection and the storage of phonetic concepts (such as stress) using the dual-coding approach;
-Provide language teachers with an editor that allows the preparation of multimedia course publishable on the Web (SMIL3.0 standard) as a template that fits the teaching of phonetic concepts. This tool must be flexible via a more user-friendly interface, and it must be as close as possible to the principle of WYSIWYG.

### 3.2 Course Models

Our proposal is that a course of phonetics must be composed with a series of lessons where each is represented by a multimedia document; so the lesson will be generated by our tool as follows:
Because each lesson is a multimedia document, its description takes into account its various dimensions:

- Structural dimension

Each lesson contains a title and a set of phonetic rules accompanied by examples. Each rule will have demonstrative examples, and each rule or example will be associated with its pronunciation (audio file). Parts of the text of an example on which the teacher wants to attract the attention of the learner are highlighted visually (wear a different color and style).

- Spatial dimension

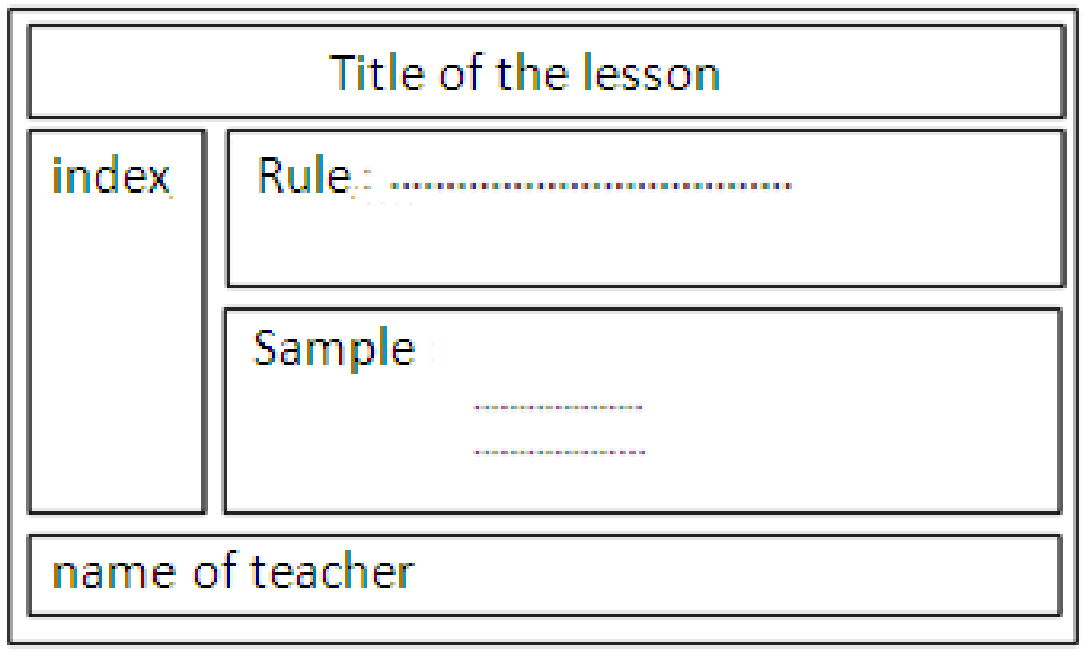

Fig. 1 Spatial dimension of a lesson.

- Temporal dimension

Each rule or example appears in parallel with its pronunciation, and they appear in sequence. The rules will follow in time one after another. The following figure shows the temporal aspect of the document.

The media objects are represented by rectangles where the length reflects the duration of display or presentation of the corresponding object.

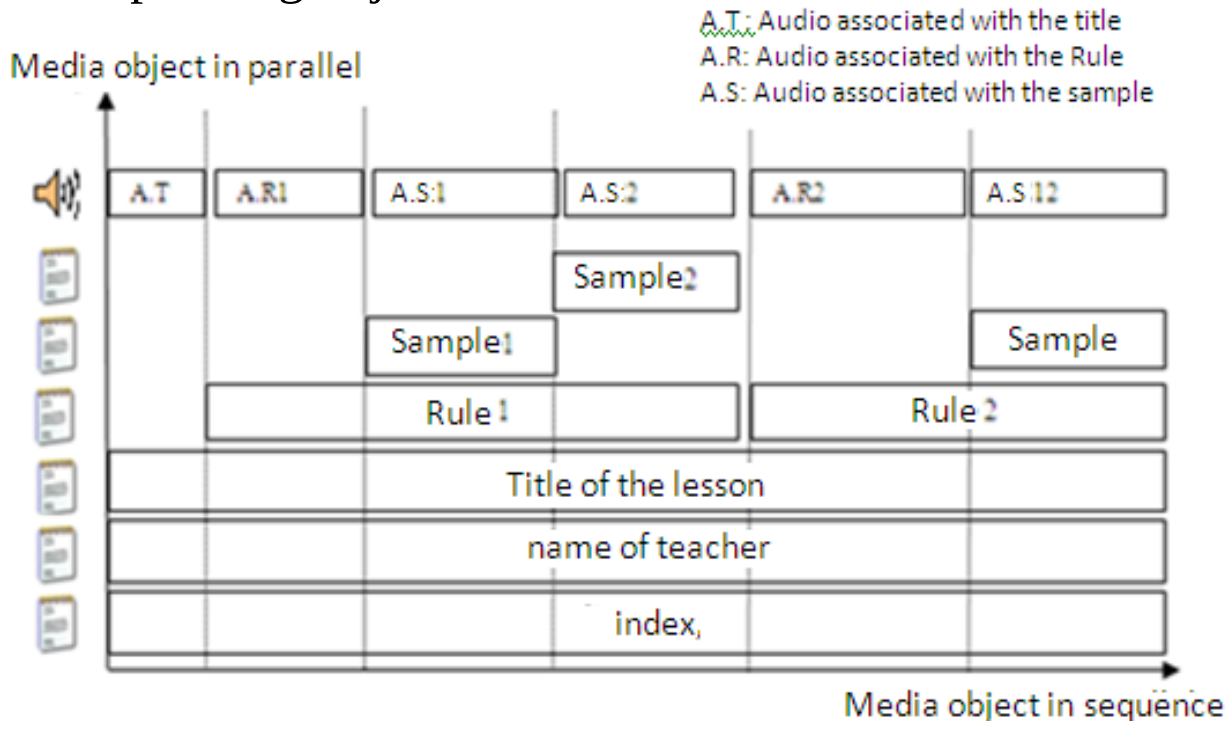

Fig. 2. Chart shows the temporal aspect of a document

- Hypermedia temporal dimension

The index object contains a list or summary of the rules of the lesson. The elements of this list are clickable areas where a click on one of them enables to watch the presentation of the lesson at the beginning of the corresponding segment (the rule in question); it represents time navigation.

## 4. Development of SACoPh System

The approach of our system is to associate a modality of typographical representation with an oral modality of accentuation. The combination of typographic style to the quality of spoken discourse has been little explored. There are three ways to combine these two types of representation: one qualified as automatic (the system Prosodic Font), one described as interactive and the last combining the two solutions (the system SWANS for example).
In our case, the solution is interactive because it allows the author (teacher) to choose its own submission on one hand and to decide where he wants to focus on the other.

### 4.1 Data Structure

The data structure of a lesson is presented by the class diagram as follows (using the UML):

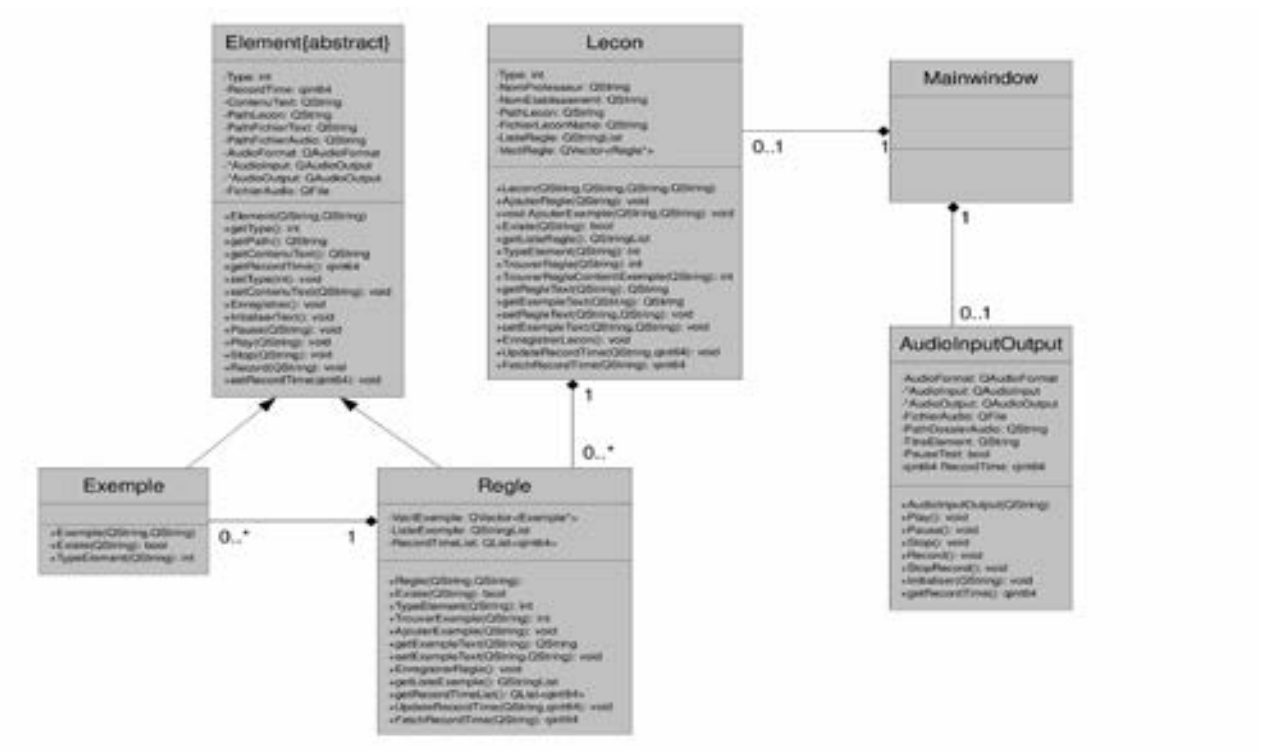

Fig. 3. Diagram of classes of a course document

## 4.2 System Architecture

The system SaCoPh consists of four modules presented in Figure 4.

Management of the data structure: This module offers to teachers the opportunity to manage the lesson by managing a tree. It can create, edit or delete a rule or an example. Deleting a rule therefore remove all the examples it contains.

Word Processing (typographical marker): Through the features of this module, the user can enter the text (of a rule or an example), insert the phonetic alphabet characters by selecting them from a list, and the most important, it can add typographic markers (color, form, style and/or size) on parts on which he wants to attract the attention of the learner. This module generates XHTML code that corresponds to such a specification for later use, to update or to facilitate the generation of SMIL code thereafter.

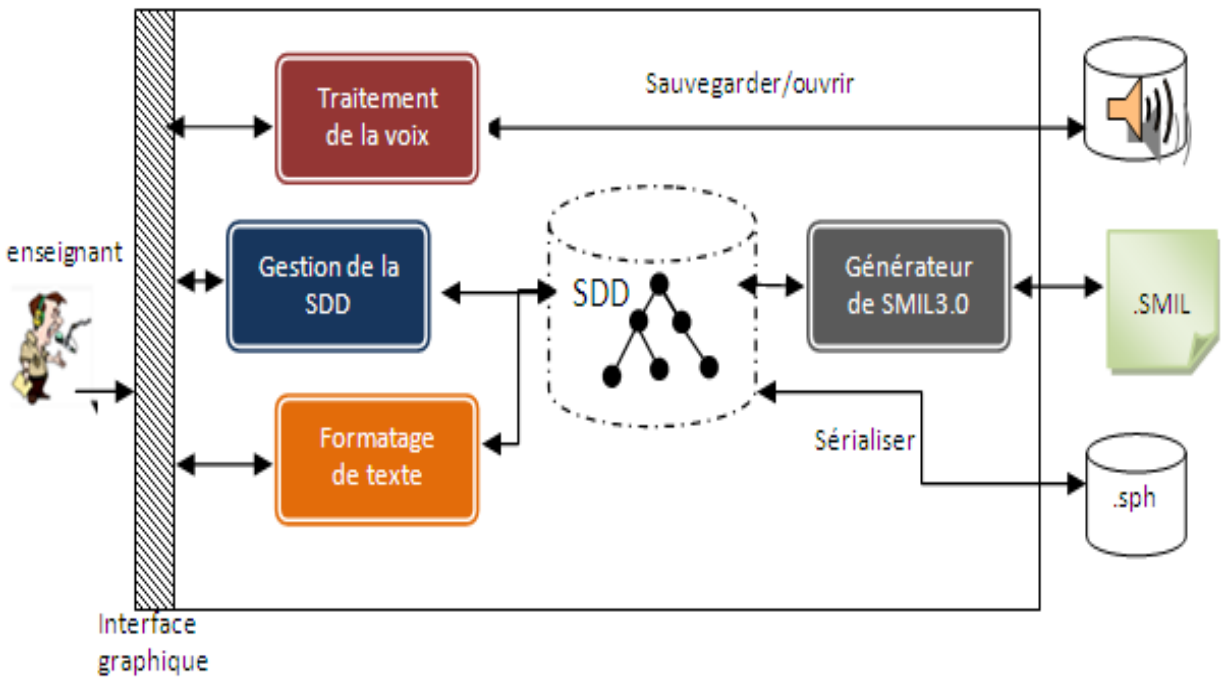

Fig. 4. SACoPh Architecture

Voice processing: This module is concerned with recording the voice of the teacher in an audio file format 'wav'. It enables to replay or to delete what was recorded and it provides information about the object as duration.

Generating of a SMIL presentation: A document published by our system is saved (serialized) into its own format (.Sph), and is ready for later updates. Nevertheless, the teacher can export documents via this module to the SMIL 3.0 format for publish or share. So far, only the player called Ambulant can read the presentations of version 3.0. Thus, this module must traverse the tree of object rules and examples (SDD) to generate the SMIL code. The synchronization between these objects is implicitly deducted from their order in the tree. The display duration of a rule or an example is deducted from the duration of the object associated.

```
<-- specify the time segment containing a rule
and its samples -->
<par xml:id="1" dur="28s">

<-- specify the pronunciation of a rule and its
samples -->

  <audio begin="1s" src="Regle 1.wav"/>
  <audio begin="11s" src="Exemple1_R1.wav"/>
  <audio begin="14s" src="Exemple2_R1.wav"/>
  …
  <-- specify a rule 1 -->
  <smilText region="Regle">
   <p>
   <span textFontFamily="…" textColor="#..."
       textFontSize="16px"> The vowel </span>
   <span textFontFamily="…" textColor="#..."
       textFontSize="18px"> a </span>
   <span textFontFamily="…" textColor="#..."
       textFontSize="16px"> is pronounced ...
       </span>
   </p>
   </smilText>

<-- specify the sample 1, witch starting in the
11th second after Rule 1 -->

   <smilText begin="11s" region="Exemple">
    <p>
    <span textFontFamily="…" textColor="#..."
        textFontSize="16px"> W </span>
    <span textFontFamily="…" textColor="#..."
        textFontSize="18px"> a </span>
    <span textFontFamily="…" textColor="#..."
        textFontSize="16px"> tch </span>

<-- specify the sample 2: witch begin at the
third second after the sample 1  -->
   <tev begin="3s"/>
   <p>
    <span textFontFamily="…" textColor="#..."
        textFontSize="16px"> B </span>
    <span textFontFamily="…" textColor="#..."
        textFontSize="18px"> a </span>
    <span textFontFamily="…" textColor="#..."
        textFontSize="16px"> th </span>
   </p>
   ...
  </smilText>
</par>
…
```

We show above some of the code generated by SMIL3.0 SACoPh to synchronize the text of a rule in parallel with pronunciation. Note that after 11s, the text of the first example is presented with its pronunciation. Note the different types of tags and attributes: those used for synchronization <par>, <tev>, begin and those used for the presentation of typographical marker <smilText>, textFontFamily, text Color, etc.

An index table is calculated automatically. We proposed a temporal segmentation of the document. This is to identify the start and duration of each segment, then place markers for each of them to be referenced later. We propose that a time segment is the time for filing a rule with all its examples: this segment is an indivisible entity. We made this choice because the rule can not be fully understood only through its explanatory examples, and they can not be divorced from the rule. Each entry in the index table, we associate a link to its corresponding time segment through its marker. The following code shows how the segments are marked, and how are referenced by entries in the index table.

```
<-- specify the sequence of temporal segments and
mark each by the tag xml :id-->
<seq>
   <-- segment 1 -->
   <par xml:id="1" dur="28s">
     <smilText region="Regle">
      ………
     </smilText>
     <smilText region="exemple">
      …
     </smilText>
      ………
   </par>
   <-- segment 2 -->
   <par xml:id="2" dur="15s">
      …………
   </par>
   ………
</seq>
  <-- complete the index table -->
   <a href="#1">
    <smilText region="Index1"> Rule 1</smilText>
   </a>
   <a href="#2">
    <smilText region="Index2"> Rule 2</smilText>
   </a>
   ………
```

The index table provides a navigation time during the show of the lesson. Students can forward or rewind the presentation to the beginning of a rule he wants replay by clicking the link in the table. Figure 5 shows the final presentation of a lesson by the ambulant player.

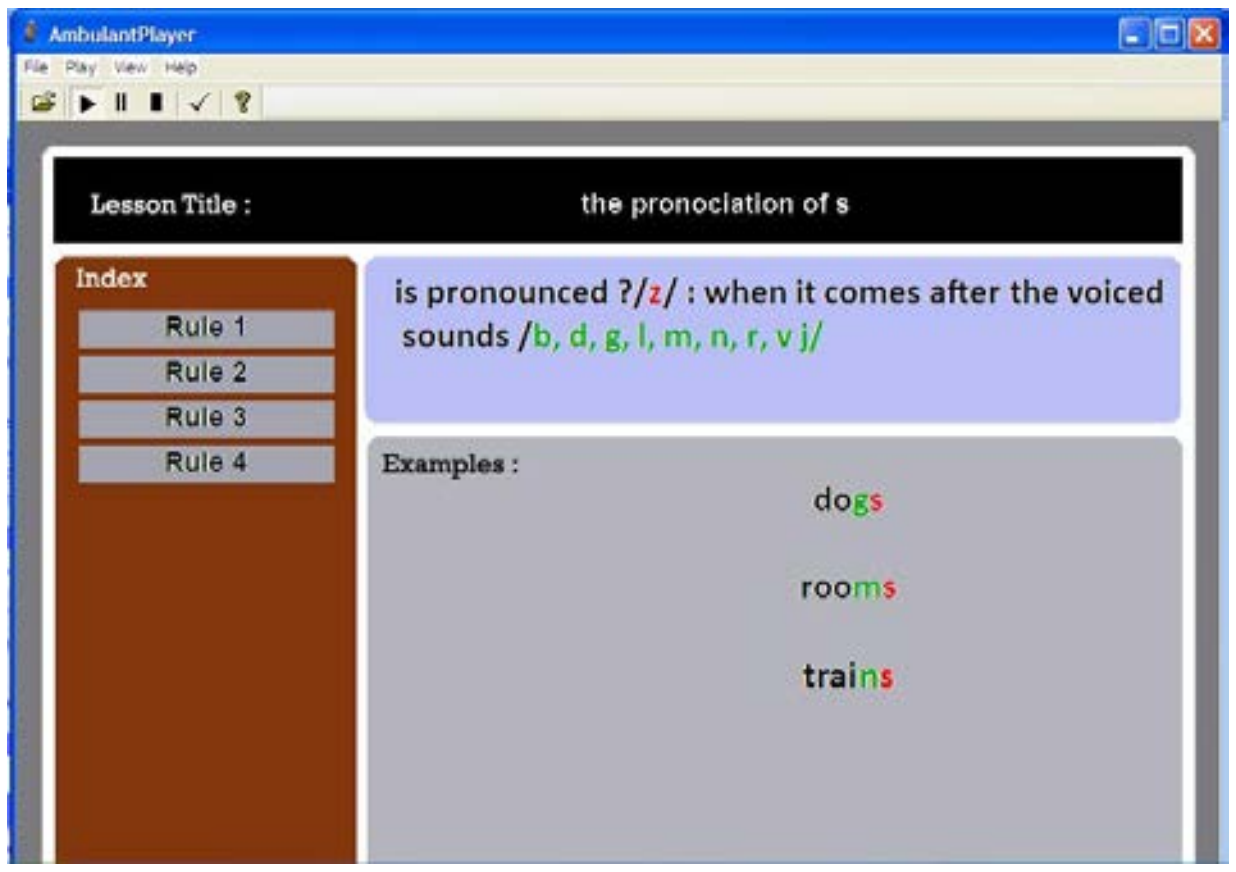

Fig. 5. – présentation d’une leçon générée avec ambulant

## 4.3 Interface SaCoPh

Our system is developed in C + + using the QT library; it allows a development of cross-platform graphical applications based on the following approach: write once and compile anywhere. All the services supported by our editor are provided via a graphical interface; it is easy to use and it takes into account the already entrenched attitudes among teachers. The figure below illustrates an overview of this interface.

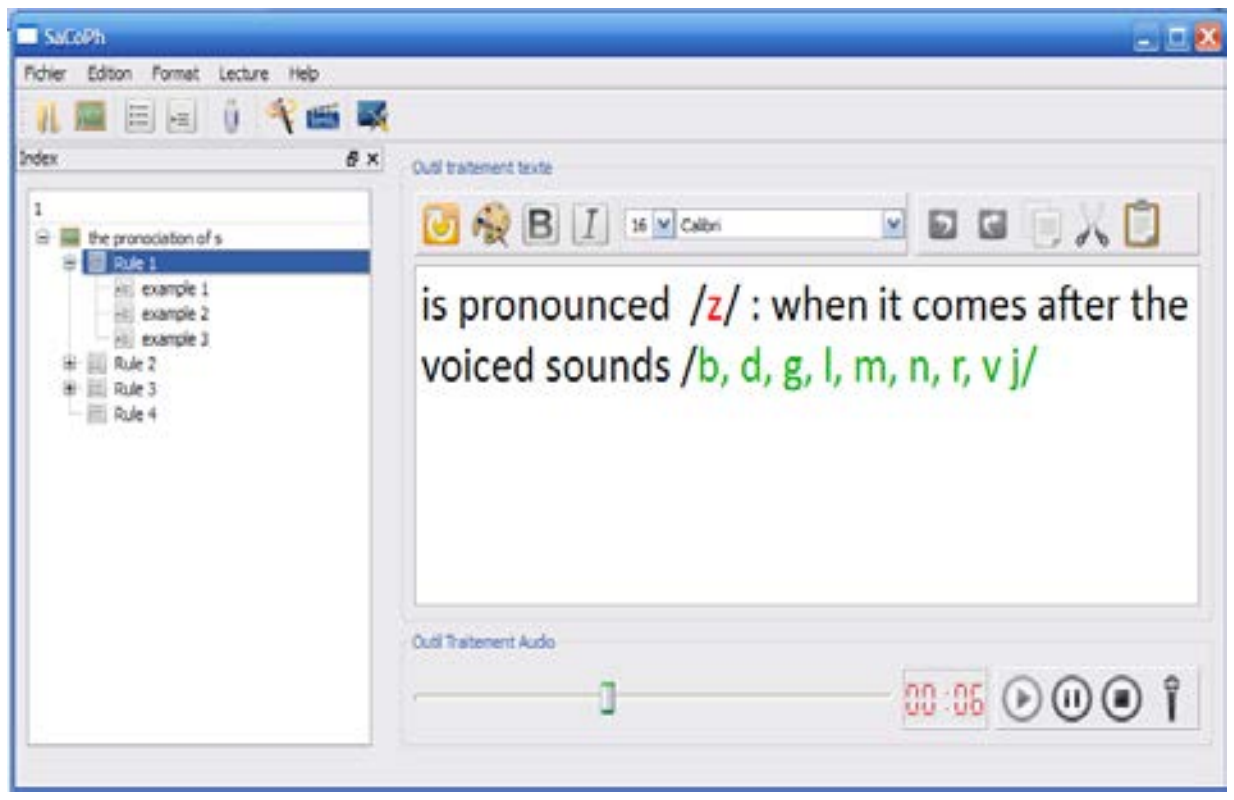

Fig. 6. GUI of SACoPh

# 5. Conclusions

Our research focus is based on the processing of multimedia documents and timed applied in distance education, and more specifically to the teaching of phonetics of a language. We presented the significant contribution of the dual coding approach in improving learning. For this, we designed a template that concretizes this approach. We have developed an authoring system called SaCoPh, which generates over phonetic publishable via the Web, according to this model, using the new features of version 3.0 of the SMIL standard. This system is designed for non-computer for teachers, it provides a simpler interface and more user-friendly as possible.
Moreover, we consider the following perspectives:
Integrate into our system functionality that allows to record video and/or draw the image. That way, the teacher will not have to use various external tools to prepare different types of media when preparing his lessons. If he wants to use an existing media, it can import the URL
Accelerate the process of publishing with a semi-automatic thinner synchronization mechanism and especially when the media is imported.
Facilitate for learners, the research via segments in lessons throughout the course of this format.

**Acknowledgments**

G. Merzougui would like to thank a lot both M. Moumni and M. Aouadj for discussions, comments and suggestions that have greatly enriched the work.

**Merzougui Ghalia** received a Master in Computer Science from the University of Batna, Algeria, in 2004. She is currently a Professor at the University of Batna, Algeria.
She is a member of (Adaptive Hypermedia in E-learning) research group. She is currently pursuing his doctoral thesis research on the management of multimedia educational content. Her current research interest is in E-Learning, system of information retrieval, ontology, semantic web, authoring and multimedia teaching resource. His teaching interests include computer architecture, software engineering and object-oriented programming, ontology and information retrieval.

**Mahieddine Djoudi** received a PhD in Computer Science from the University of Nancy, France, in 1991. He is currently an Associate Professor at the University of Poitiers, France.
He is a member of SIC (Signal, Images and Communications) Research laboratory. He is also a member of IRMA E-learning research group. His PhD thesis research was in Continuous Speech Recognition. His current research interest is in E-Learning, Mobile Learning, Computer Supported Cooperative Work and Information Literacy. His teaching interests include Programming, Data Bases, Artificial Intelligence and Information & Communication Technology. He started and is involved in many research projects which include many researchers from different Algerian universities..